# Photonic single perceptron at Giga-OP/s speeds with Kerr microcombs for scalable optical neural networks

Mengxi Tan[1], Xingyuan Xu[2], and David J. Moss[1]

[1]*Optical Sciences Centre, Swinburne University of Technology, Hawthorn, VIC 3122, Australia*
[2]*Dept. of Electrical and Computer Systems Engineering, Monash University, Clayton, 3800 VIC, Australia*

**Optical artificial neural networks (ONNs) have significant potential for ultra-high computing speed and energy efficiency. We report a novel approach to ONNs that uses integrated Kerr optical micro-combs. This approach is programmable and scalable and is capable of reaching ultra-high speeds. We demonstrate the basic building block ONNs — a single neuron perceptron — by mapping synapses onto 49 wavelengths to achieve an operating speed of 11.9 x $10^9$ operations per second, or Giga-OPS, at 8 bits per operation, which equates to 95.2 gigabits/s (Gbps). We test the perceptron on handwritten-digit recognition and cancer-cell detection — achieving over 90% and 85% accuracy, respectively. By scaling the perceptron to a deep learning network using off-the-shelf telecom technology we can achieve high throughput operation for matrix multiplication for real-time massive data processing.**

## I. Introduction

Artificial neural networks (ANNs) have achieved significant success in making predictions and achieving simple representations of complex and high dimension data. When sufficient data are used for training, ANNs can outperform computational algorithms [1-5] and even humans for many tasks ranging from the recognition of images to translation of languages, risk assessment and intriguingly, complex board games [6]. The speed and computational power of ANNs is determined by matrix multiplication operations, or multipy-and-accumulate (MAC) operations. Electronic ANN chips include the IBM TrueNorth and Google TPU chips [7, 8]. They use extremely large-scale processor arrays that include the systolic array [8], to enhance the parallelism to achieve operational speeds greater than 180 x $10^{12}$ floating point operations per second (Tera-FLOPS). However, in spite of this performance, since they are electronically based they are still subject to relatively inefficient digital protocols and bandwidth bottlenecks due to the von Neuman effect [9]. In fact, each individual processor is limited in speed to only about 700 MHz [10 9].

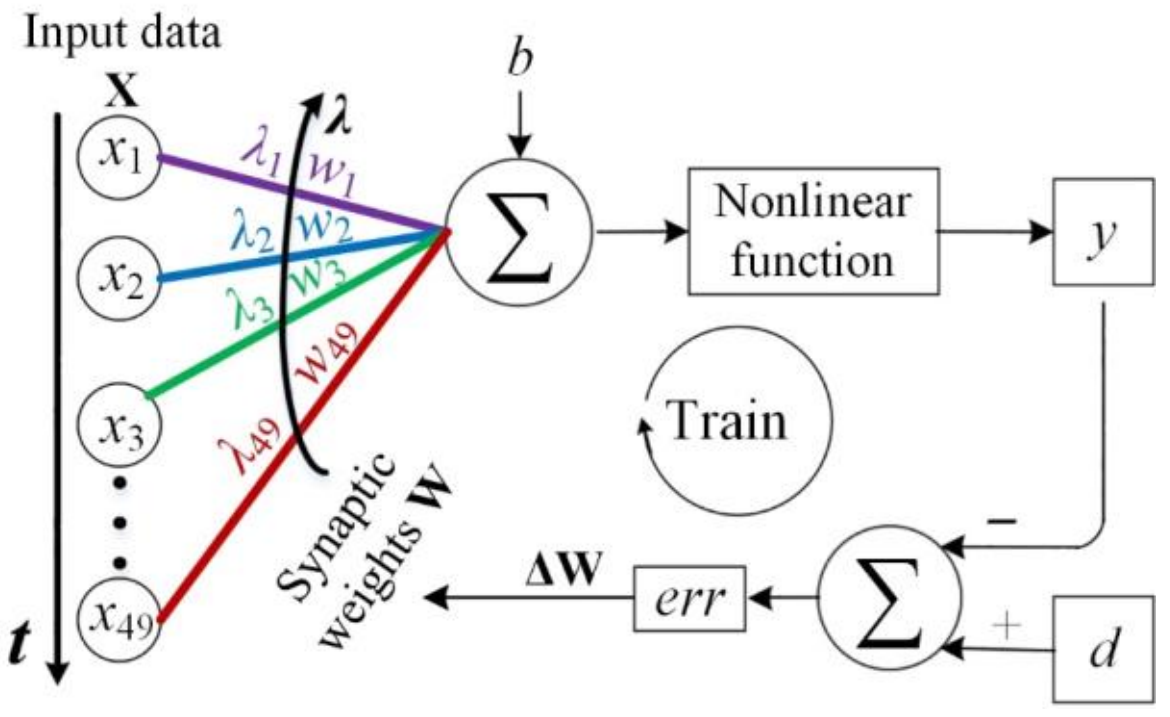

Figure 1. Mathematical model of perceptron.

Photonic approaches towards ANNs, or optical neural networks (ONNs), are next generation neuromorphic processors and are attracting extremely high levels of interest currently [11-17]. They are highly promising since they offer the potential to achieve extremely high processing speeds [3]. The key is to achieve the weighted synapses that connect the nodes and neurons. In contrast to electronic digital systems that store the synapses in memory, photonic systems operate by realizing actual physical embodiments of synapses, where their number determines the scale of the network, and which depend on the physical parallelism that is fundamentally analog in nature.

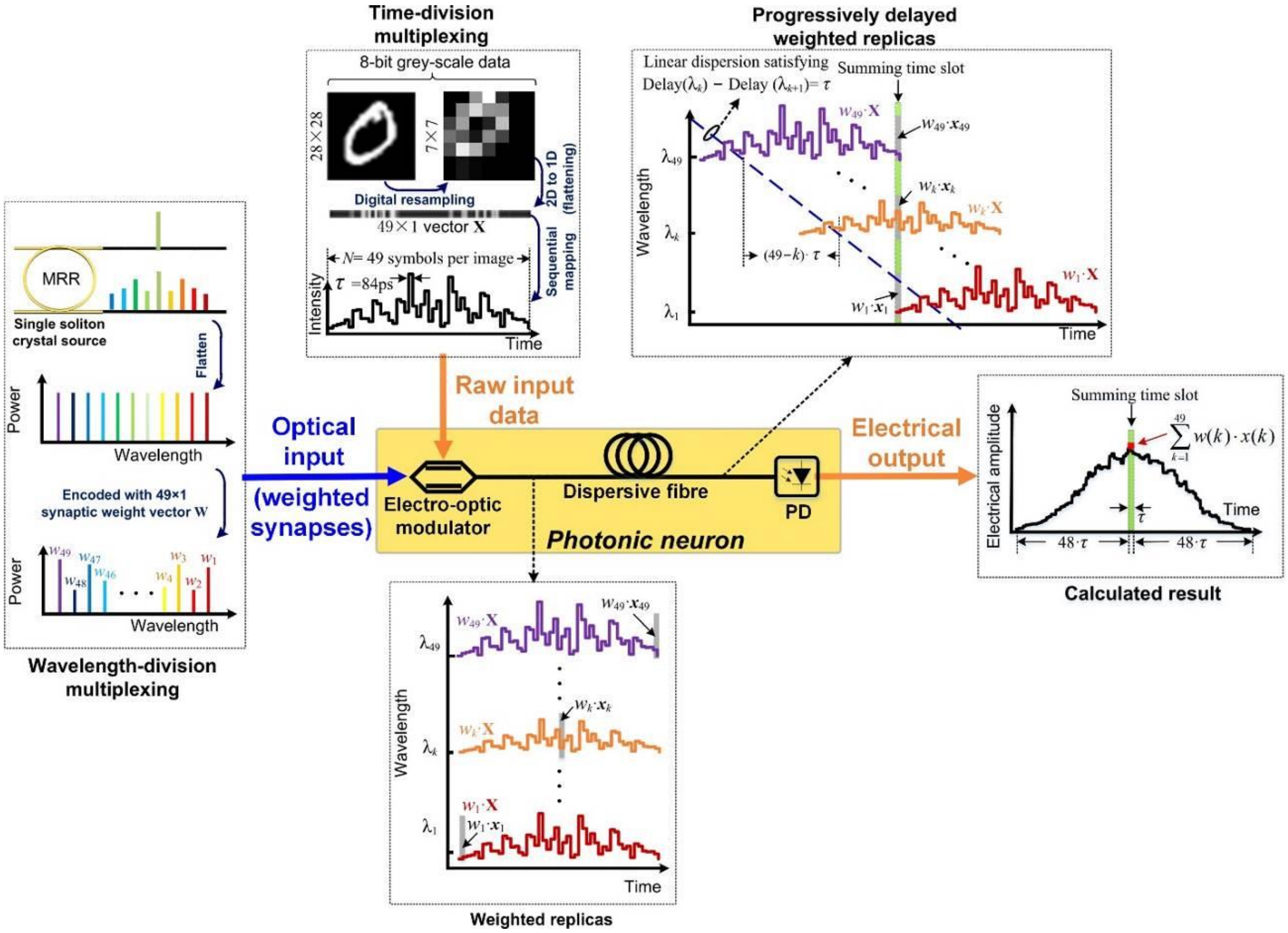

Figure 2. Experimental setup for single perceptron.

ONNs have achieved substantial success by using a number of approaches to multiplex synapses in parallel. Schemes based on spatial multiplexing include coherent integrated photonic chips [3] and diffractive bulk optics [17]. These have successfully achieved classification of alphabetic-numeric characters including handwritten digits and vowels. Furthermore, they have achieved low power levels, although they have tradeoffs between the system footprint and processing power, determined by the number of synapses or degree of parallelism. There are a number of ways to realise ONNs, including reservoir computing [18-21] as well as spike processing [22-25]. These both use sophisticated schemes to multiplex the synapses and are both very compact. Photonic reservoir computing multiplexes the synapses in time to achieve very large scale systems with hundreds of input layer nodes. Conversely, spike processors have successfully achieved pattern recognition by using phase change materials in integrated devices [24]. This approach operates via wavelength division multiplexing (WDM) and also benefits from a dynamically reconfigurable operation bandwidth [17-19]. Despite the success of these approaches, they still face limitations. Temporal multiplexing is challenging to dynamically train and to scale up to deep learning systems with multiple layers. Spike processing is limited in the degree of parallelism it can achieve because it relies on arrays of discrete laser diodes. The combined use of temporal, wavelength, and spatial multiplexing has the greatest potential to achieve the highest combination of processing power, operation speed and scale of the network, and this is what our approach uses.

## II. PERCEPTRON

Here, we propose [11, 12] a novel scheme for ONNs based on integrated micro-combs to achieve simultaneous temporal, spatial, and wavelength multiplexing, which we then use to perform the dot product of vectors. We perform matrix operations by first flattening the matrices to convert them into vectors at high data rates. Our system is capable of dynamic training and its network structure is highly scalable. We demonstrate a single photonic neuron perceptron with 49 synapses, or wavelengths using the microcomb. Our fundamental building block for ONNs achieves a speed for matrix multiplication at 11.9 billion (Giga) operations/s (OPS) – or GOPs - that equates to 95.2 Gigabits/s for 8 bit operations. We do this via simultaneous synapse weighting in the wavelength domain and the temporal domain, scaling the input data. The device is applied to benchmark tests that include handwritten digit classification, where we obtain an accuracy greater than 93%, and to the prediction of cancer classes to distinguish malignant from benign cases based on an extracted feature set from microscope images from biopsied tissue. We obtain an accuracy of greater than 85% for cancer classification.

Figure 1 depicts the neuron perceptron mathematical model [25] and Figure 2 outlines the experiment setup that uses a Kerr optical micro-comb source. The perceptron is based on wavelength multiplexing of 49 microcomb wavelengths, done simultaneously with temporal multiplexing, in order to form a single synapse. The main operation consists of matrix multiplication with vectors formed from flattened matrices. The matrix multiplication occurs between the electronic image input data and the synaptic weights, and this is performed with multiple steps using photonics. The input data for classification consists of $28 \times 28$ electronic digital matrices with 8-bit grey-scale intensity resolution, which is initially down sampled digitally into 7×7 matrices that are then reorganized into 1D vectors: X(i) = [X(1), X(2) … X(49)], which are then multiplexed sequentially in the temporal domain by an electronic high speed D/A converter at 11.9 Gigabaud. Here, each symbol corresponds to the 8-bit pixel input data images and takes up one time slot 84 ps in length. Hence, the whole duration of the waveform is $N \times \tau = 4.12$ ns with N=49. In approaches based on digital electronics, the neural network input nodes usually reside in electrical memory and are routed according to memory address. By comparison, the input nodes for the ONN are temporally defined by multiplexing the symbols that are then routed according to their location in time.

Following this, the electrical input waveform that is a temporally multiplexed signal is broadcast via an electro-optic modulator on to all 49 wavelengths (equal to the number of elements of the vector X), the wavelengths being generated by the micro-comb. Here, each comb line contains an equal copy of X, the time domain multiplexed input data waveform. Every comb line's power is then adjusted by an optical waveshaper with the weights being determined by the theoretical synaptic weight vector W = [w(1), w(2), …, w(49)] obtained during training. This effectively multiplexes the synaptic weights in wavelength. If W and X are both 1×49 column vectors, then the weighted input X vector replicas are

$$\mathbf{X}\times\mathbf{W}^T=\begin{pmatrix} w(1)\cdot x(1) & w(1)\cdot x(2) & w(1)\cdot x(3) & \cdots & w(1)\cdot x(49) \\ w(2)\cdot x(1) & w(2)\cdot x(2) & w(2)\cdot x(3) & \cdots & w(2)\cdot x(49) \\ w(3)\cdot x(1) & w(3)\cdot x(2) & w(3)\cdot x(3) & \cdots & w(3)\cdot x(49) \\ \vdots & \vdots & \vdots & \ddots & \vdots \\ w(49)\cdot x(1) & w(49)\cdot x(2) & w(49)\cdot x(3) & \cdots & w(49)\cdot x(49) \end{pmatrix} \tag{1}$$

where the nth row (where $n \in (1, N)$) corresponds to the temporal weighted waveform replica of the $n^{th}$ wavelength. Therefore, the diagonal components reflect the input N weighted nodes, so that the $n^{th}$ weighted input node is reflected in the 8-bit symbol w(n)·x(n) that exists in the $n^{th}$ time slot for the $n^{th}$ wavelength. After this, the replicas are transmitted through a medium that provides a dispersive delay equivalent to $2^{nd}$ order dispersion, to sequentially delay the weighted replicas in order to align the diagonal components into the same time window, with the delay step given by $\tau = \mathrm{delay}(\lambda_k) - \mathrm{delay}(\lambda_{k+1})$. Therefore, the dispersive delay is an addressable time-of-flight memory that lines up the progressively weighted time-dependent symbols w(1) · x(1), w(2) · x(2) … w(49) · x(49) over all wavelengths as

$$\begin{pmatrix} & & & & w(1)\cdot x(1) & \cdots & w(1)\cdot x(47) & w(1)\cdot x(48) & w(1)\cdot x(49) \\ & & & w(2)\cdot x(1) & w(2)\cdot x(2) & \cdots & w(2)\cdot x(48) & w(2)\cdot x(49) & \\ & & w(3)\cdot x(1) & w(3)\cdot x(2) & w(3)\cdot x(3) & \cdots & w(3)\cdot x(49) & & \\ & \iddots & \vdots & \vdots & \vdots & \iddots & & & \\ w(49)\cdot x(1) & \cdots & w(49)\cdot x(47) & w(49)\cdot x(48) & w(49)\cdot x(49) & & & & \end{pmatrix} \tag{2}$$

While this process, as it is implemented here, does not enhance the network speed because it only uses diagonal components, in principle a significant increase in speed can be obtained by scaling the network to deep (multiple level) structures through the use of parallel wavelengths as well as time and spatial multiplexing.

Finally, the intensity of all of the optical signals in each time bin are summed via sampling and detection to produce the resulting matrix multiplication (equivalent to a dot product of 49×1 vectors for the case of 7×7 matrices) of the neuron, given by:

$$\mathbf{X}\cdot\mathbf{W}=\sum_{k=1}^{49} w(k)\cdot x(k) \qquad (3)$$

After matrix multiplication, the summed, weighted output is then modulated in order to map it into a desired range by using a nonlinear sigmoid function. In this initial demonstration we achieve this last function using digital electronics, which generates the output of the single neuron perceptron. In principle, however, this can easily be achieved all-optically. Finally, the input data prediction category is produced through comparison between the decision boundary with the neuron output. The decision boundary is a 49 dimensional hyperplane, generated during digital learning carried out offline prior to the experiments. Thus, the input data can be separated into two categories.

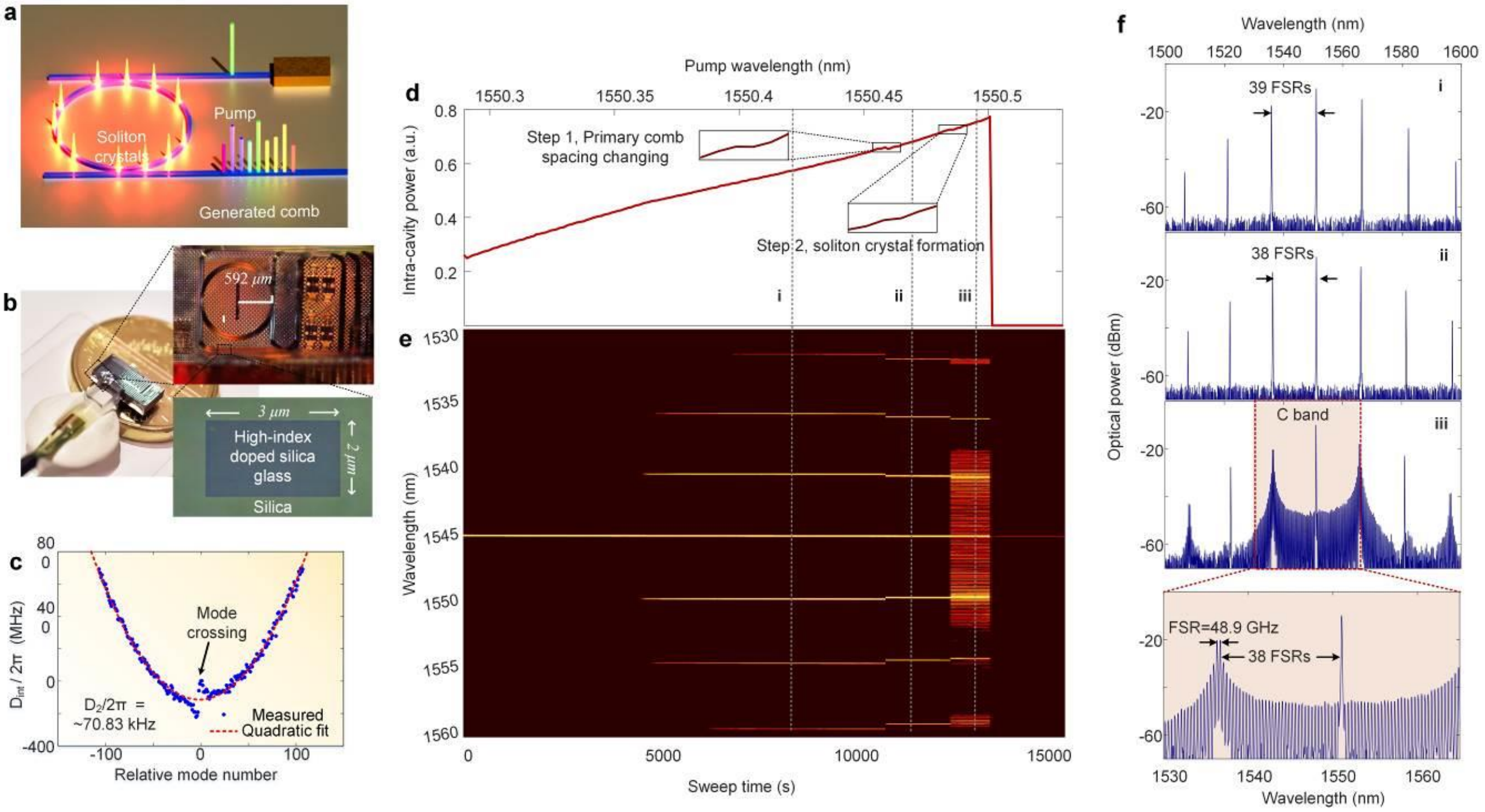

**Figure 3 | a.** Schematic diagram of the soliton crystal microcomb, generated by pumping an on-chip high-Q nonlinear micro-ring resonator with a continuous-wave laser. **b.** Image of the MRR and a scanning electron microscope image of the MRR's waveguide cross section. **c.** Measured dispersion of the MRR showing the mode crossing at ~1552 nm. **d.** Measured soliton crystal step of the intra-cavity power, and **e.** optical spectrum of the microcomb when sweeping the pump wavelength. **f.** Optical spectrum of the generated coherent microcomb at different pump detunings at a fixed power.

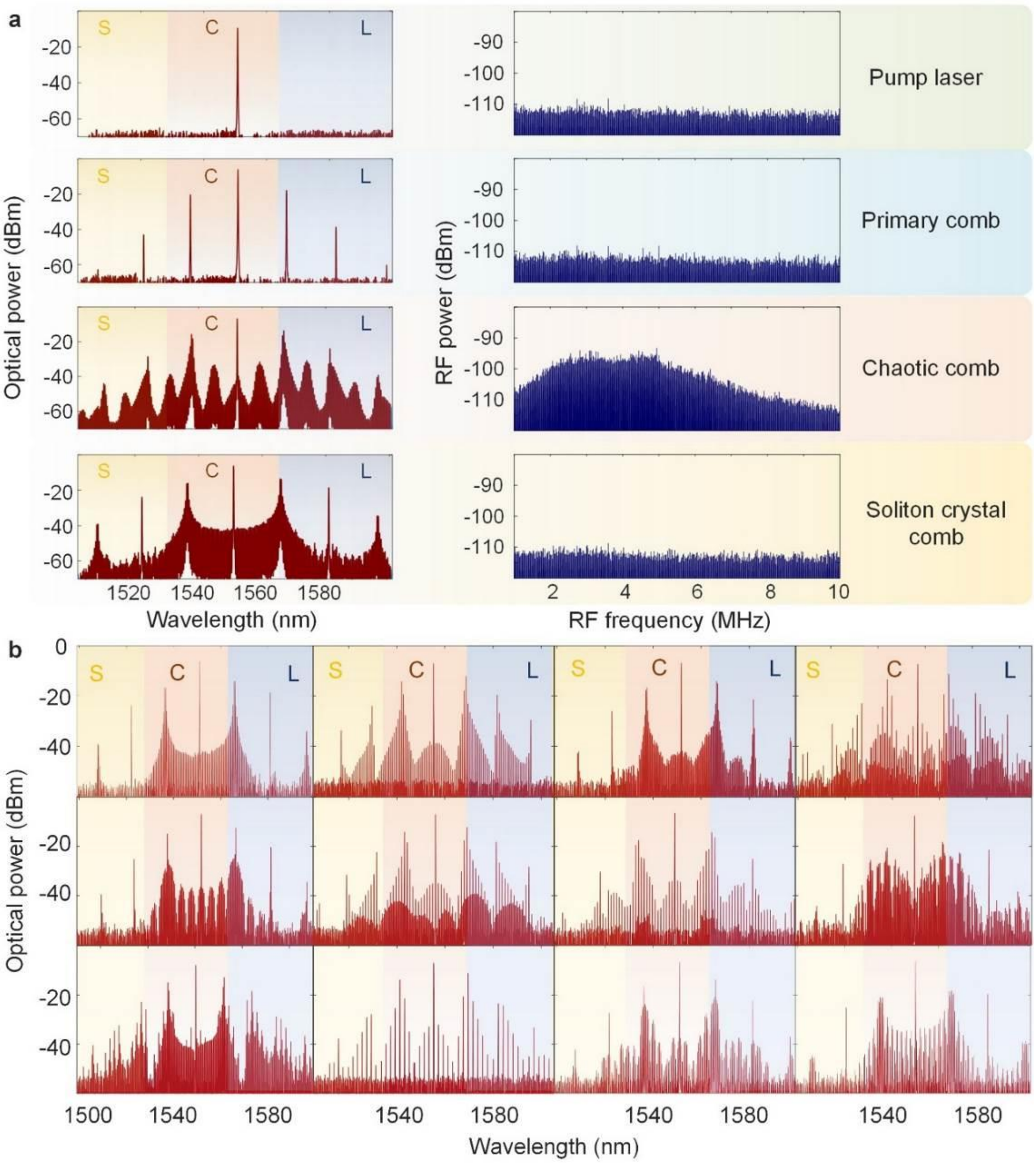

**Figure 4 | a,** Different states and measured RF intensity noise of the microcomb. **b,** Measured low intensity noise states.

## III. MICRO-COMBS BASED ON SOLITON CRYSTALS

Kerr optical micro-combs [26-33] have achieved many ground-breaking breakthroughs including optical frequency synthesis [29], ultra-high bit rate data transmission [30], generation of advanced quantum states [31], high level RF signal processing [32], and more. They provide the full capability of mainframe optical frequency combs [34] although in a fully integrated form that has a much more compact footprint as well as the potential to scale the network in power, reliability, and performance [35-42]. The new platforms developed for optical microcombs [27,43,44] have much lower nonlinear absorption than other nonlinear platforms such as chalcogenide glass and semiconductors [45-67].

We use a microcomb that operates via soliton crystal states [68, 69], produced in integrated ring resonators. Soliton crystals display deterministic generation induced by mode crossings that produce a background wave, with all of these processes sustained by the Kerr nonlinearity. For soliton crystals, there is very little nonlinear pump-induced shift in the resonance that otherwise would require difficult dynamic pumping schemes like DKS solitons require [26]. This is because the intracavity soliton crystal state power is virtually the same as the power for the chaotic state from which it is formed. Hence, very little power jump occurs when they are generated and this allows a reliable and simple method of initiation achievable by simple adiabatic, even manual, tuning of the pump wavelength [68]. This same effect also yields a much higher energy conversion efficiency from pump to comb-line [63]. Soliton crystals have demonstrated a multitude of RF or microwave signal processing based on photonics [11,12,32,70-94]. The integrated ring resonators were made from Hydex glass, a platform that is CMOS compatible [27] (Figure 3). They had a high Q factor of 1.5 million and a 48.9 GHz FSR with a chip to fibre coupling loss of 0.5 dB / facet achieved by on-chip mode converters. The waveguide cross-section of 3μm × 2μm produced anomalous dispersion with a mode crossing near 1552 nm. A 30dBm CW pump laser generated the soliton crystals when its wavelength was swept manually from short to long wavelengths (blue to red) near a resonance.

To generate coherent micro-combs, a CW pump laser (Yenista Tunics – 100S-HP) was employed, with the power amplified to 30dBm by an optical amplifier (Pritel PMFA-37) and the wavelength subsequently swept from blue to red. The acquired soliton crystals optical spectra are shown in Figures 3 and 4. We note that when locking the pump wavelength to the resonance of the MRR, the stability of the microcomb can be further enhanced that could even serve as frequency standards [29]. Figures 3f and 4a show the progression from the initial onset of primary combs, to chaotic combs that are not modelocked, to finally soliton crystal combs. Also shown in Figure 4b are the range of different soliton crystal states that can be obtained by adjusting the pump offset (to the nearest resonance) as well as the overall pumping wavelength. Figure 3 also indicates that the power jump in transitioning from the chaos state to soliton crystal state is extremely small. This arises because the two states have very similar power levels. This is a key reason for the stability of soliton crystals.

## IV. EXPERIMENT

As discussed above, the multicasting of the waveform was achieved via intensity modulation of all of the wavelength channels supplied by the shaped comb lines, simultaneously. Hence, the optical signal at the $k$th ($k$=1, 2, …, 49) channel was $w(k)\cdot\mathbf{X}$. The delay that we used for the optical signals at all wavelength channels was 13-km of dispersive single mode fibre which generated a time delay of $(49 - k)\times\tau$ for the $k$th channel, and $\tau$ was measured to be 84 ps. Thus, the optical signals were progressively shifted in the time domain. The optical signal after the single mode fibre was converted to the electrical domain by a photodetector (Finisar VPDV2120), and the waveform was then measured by a high-speed oscilloscope (Keysight DSOZ504A). The sampled output of the photodetector was added to the bias symbol and rescaled in intensity by the reference symbol to extract the recovered output of the ONN and locate the hyper-plane (a trained subspace in the high-dimension space of the input data, which serves as a decision boundary that separates different classes of data).

During the experiment, the 7×7 gray scale data of the handwritten digit figures were first converted into a one dimension array $\mathbf{X}=[x(1), x(2), \ldots, x(49)]$ by assembling each column head-to-tail. Then a 49-symbol waveform was generated and coded with the intensities at each time slot in proportion to the values of **X** at corresponding sequences, thus the input data **X** were multiplexed in the time domain. The 49-symbol waveform was generated by an arbitrary waveform generator (Keysight M8195A), which supported a sample rate of 65 Giga-Samples/s and an analog bandwidth of up to 25 GHz. To acquire stepwise waveforms for the input nodes, we used 5 sample points at 59.421642 Giga Samples/s to form each single symbol of the input waveform, which also matched the progressive time delay $\tau$ (84 ps) of the dispersive buffer.

The optical power of the 49 microcomb lines was shaped according to the intensity of pre-trained neuron weights $\mathbf{W}=[w(1), w(2), \ldots, w(49)]$. We shaped the comb lines' power with a programmable optical spectral shaper using liquid crystal on silicon techniques (Finisar WaveShaper 4000S), which could dynamically reconfigure the ONN connections within 500 ms with a resolution of 1 GHz. Two stages of programmable optical spectral shapers were employed for a larger loss dynamic range. The first WaveShaper was used to flatten the microcomb, while the second one was used to achieve pre-trained neuron weights. A feedback loop was used to enhance the shaping accuracy, where the comb lines' power after shaping was measured by an optical spectrum analyser (Anritsu MS9710C) and compared with the pre-trained weights to generate an error signal for the calibration of the WaveShapers' loss characteristics.

Figure 5 shows the time-domain multiplexed input layer for the cancer diagnosis test. The generated 11.9 Giga-baud data stream of the encoded 75 sets of features shows the 30-symbol encoded data for each set and 3 symbols padded for post measurement, including a trigger symbol to trigger the oscilloscope, a reference symbol to calibrate the reference level, and a bias symbol encoded with the pre-trained bias to locate the decision boundary. Figure 6 shows the experimental recognition of cancer diagnosis. Figure 6a shows the optical spectrum of the shaped (soliton crystal) micro-comb measured by an optical spectrum analyser, while Figure 6b shows the measured and sampled output waveform from the photodetector. Figure 6c shows the recovered ONN predictions X×W+b acquired by rescaling the sampled results via the reference symbol, and the hyper-plane X×W+b=0 (black line).

## V. DATASETS AND PRE-TRAINING

The datasets we employed was from MNIST (Modified National Institute of Standards and Technology) handwritten digit database [95] and part of the publicly available Wisconsin Breast Cancer dataset [96]. The datasets of recognition tasks were first separated as training sets and test sets. The training sets were used for the offline training with the Back Propagation algorithm [97], performed on an electronic computer using Matlab™, to acquire pre-trained weights and bias. The test sets were tested with both the ONN and an electronic computer for comparison. We note that, since the number of training sets is sufficiently large compared with the number of synaptic connections, the cross validation process was not employed in this work—and in any case, it could be performed offline before the pre-training.

We note here that the accuracy of the ONN predictions was experimentally limited by the performance of the arbitrary waveform generator, which introduced errors to the symbols' intensities and thus deteriorated the correctness of the matrix multiplication. This can be addressed by using an arbitrary waveform generator with a larger analog bandwidth, or a higher sampling rate. Addressing this issue would result in higher levels of correctness than reported here.

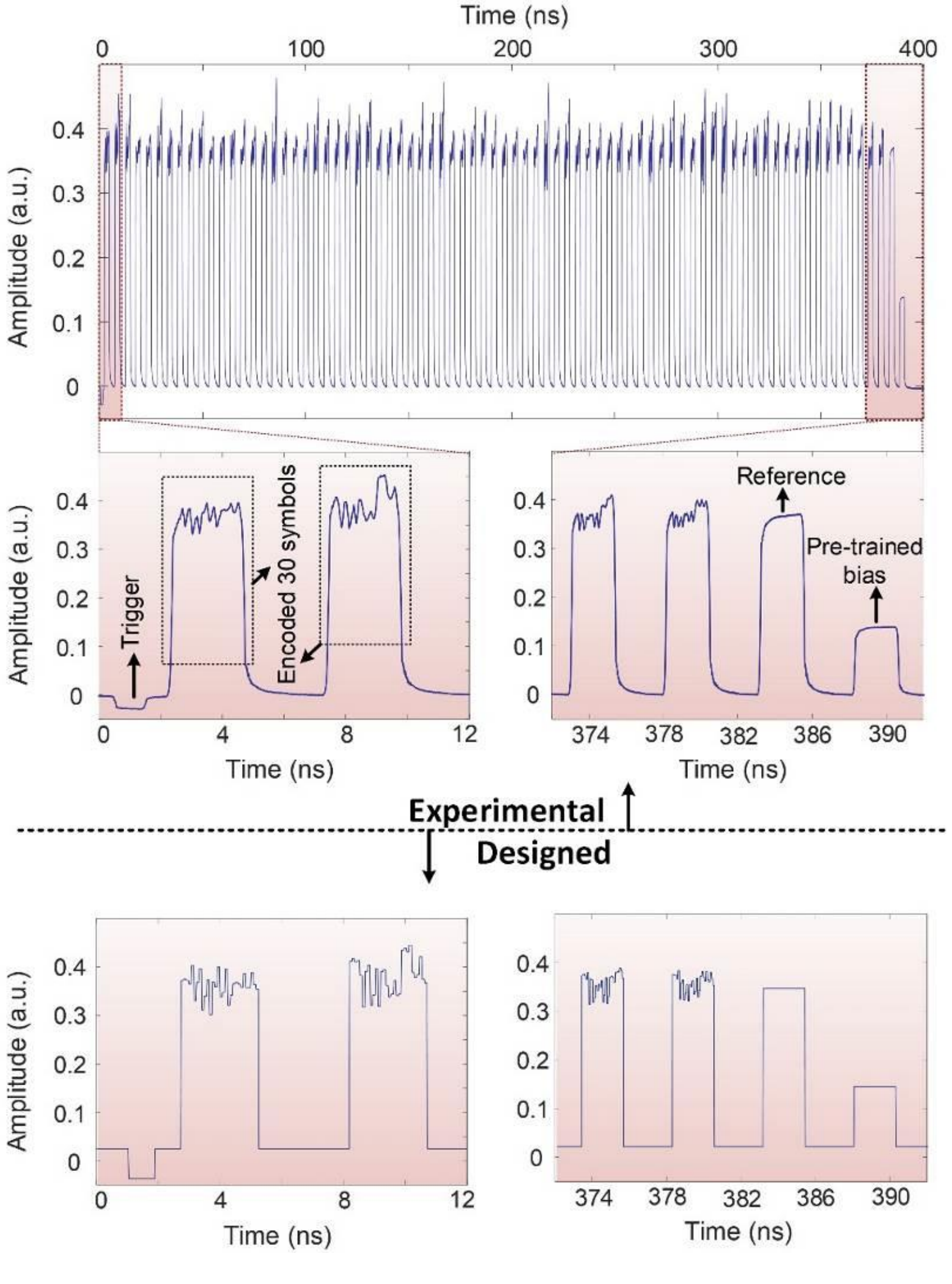

**Figure 5** | Time-domain multiplexed input layer of cancer diagnosis test. Generated 11.9 Giga-baud data stream of the encoded 75 sets of features showing 30-symbol encoded data for each set and 3 symbols padded for post measurement, including a trigger symbol to trigger the oscilloscope, a reference symbol to calibrate the reference level, and a bias symbol encoded with the pre-trained bias to locate the decision boundary.

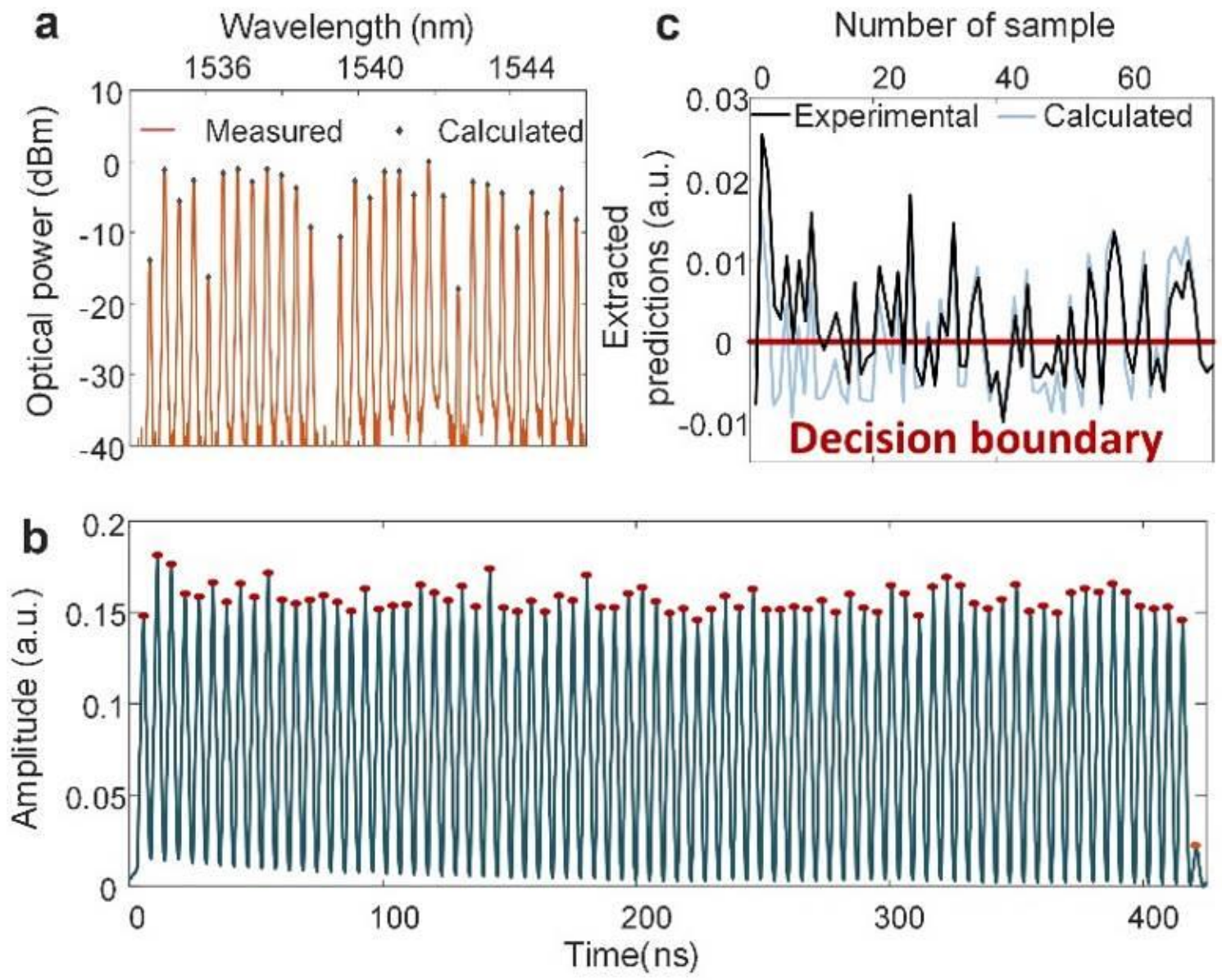

**Figure 6 | Experimental recognition of cancer diagnosis. a,** Optical spectrum of the shaped (soliton crystal) micro-comb measured by an optical spectrum analyser. **b,** Measured and sampled output waveform from the photodetector. **c,** Recovered ONN predictions X×W+b acquired by rescaling the sampled results via the reference symbol, and the hyper-plane X×W+b=0 (black line).

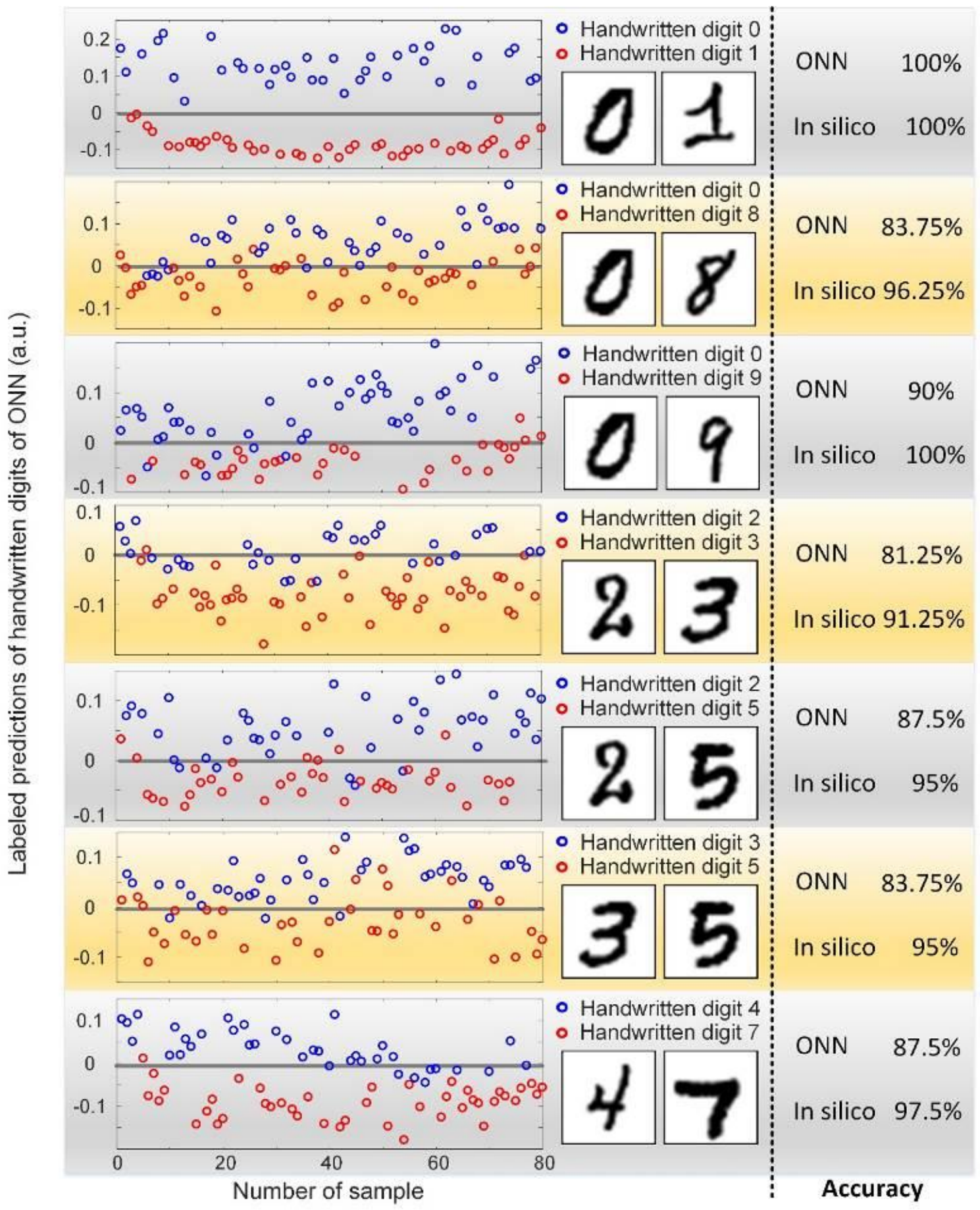

**Figure 7 S2 |** ONN predictions of handwritten digits labeled according to their correct answers.

# VI. Results

First, we evaluated the performance of the network using a number of handwritten digit pairs from a body of 500 images for each digit, from which we randomly selected 920 images for prior off-line training, which left 80 figures to evaluate the system performance. The handwritten digital images were electronically down-sampled to reduce the size of the images to 7×7 from 28×28. Next, this was transformed into a 49 symbol one dimensional array, following which the array was temporally multiplexed with each symbol occupying an 84ps time slot, yielding a modulation rate of 11.9 Gigabaud. The data vector dimension of our perceptron needed to match the weight vector dimension, given by number of wavelength, which was 49. Therefore, we used a down-sampling method on the image to reduce the length of the vector to 49.

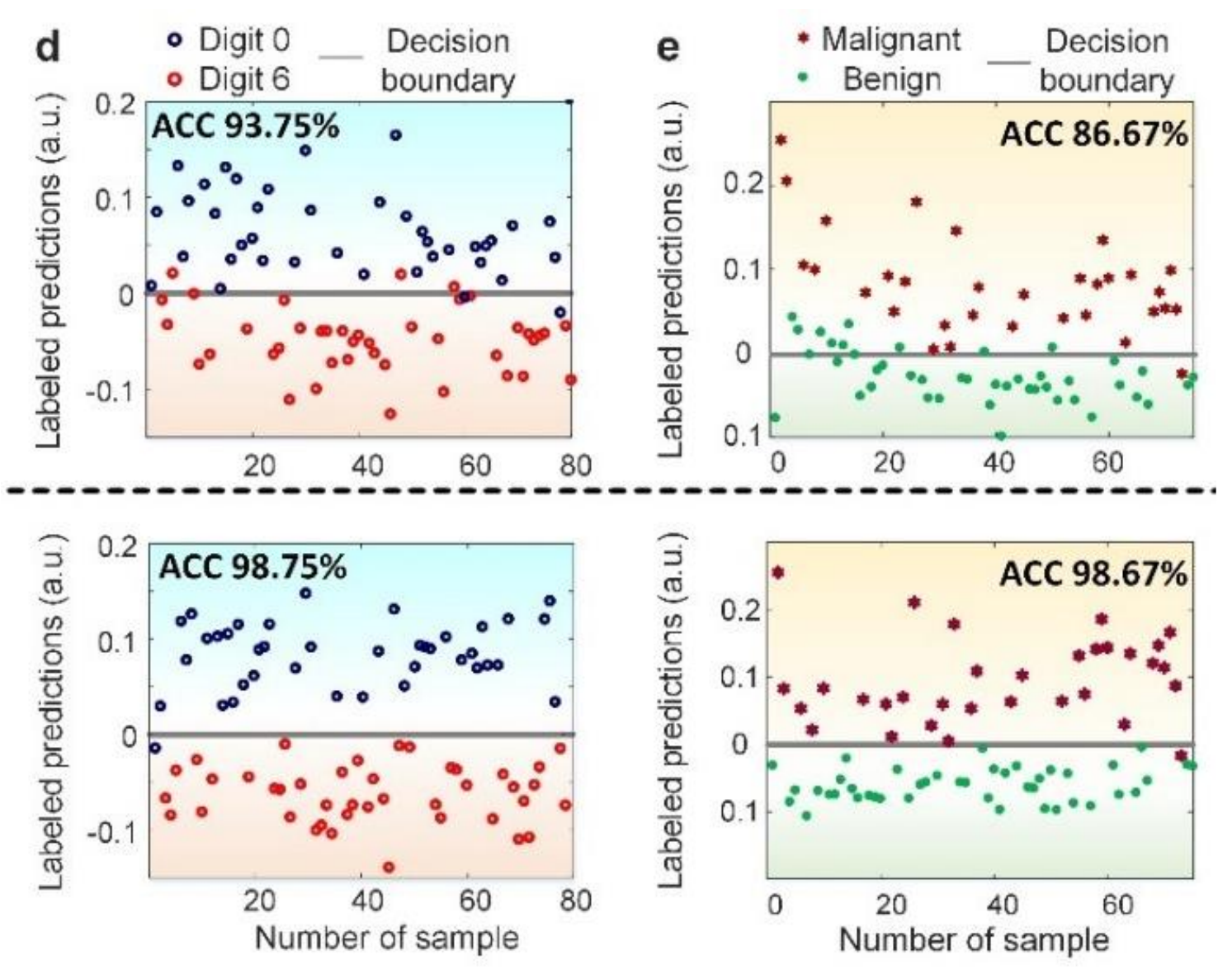

Figure 8 3. Results. d, Predictions for handwritten digit recognition showing an accuracy of 93.75% vs 98.75% for theory. e, Predictions for benign versus malignant tumor cells with an accuracy of 86.67%.

The optical power for each of the 49 comb lines was weighted according to the pre-learned synaptic weights in order to enhance the parallelism to form the neuron synapses. Next, the data input stream was simultaneously imprinted onto all of the 49 weighted microcomb lines, which were then linearly progressively delayed in wavelength by 13km of single mode fibre that generated a time-of-flight optical buffer via its 2$^{nd}$ order dispersion of 17 ps / nm / km. Therefore, the weighted symbols for each wavelength were aligned in time, thereby enabling them to be summed by simple sampling of the centre timeslot and subsequent detection. This yielded the matrix multiplication result, a product of the multiply and accumulate (MAC) operation. The output was finally compared against the decision boundary which consists of a hyper-plane that was generated during prior network training that classified the input samples arranged in a 49-dimensional hyperspace. The resulting matrix multiplication computations on the multiple input data samples were then compared in intensity against this decision boundary, finally producing the predictions of the perceptron (Figures 7, 8). We tested the perceptron performance for classifying 2 benchmark tests delineated by the decision boundary – first for two handwritten digits (0 and 6), followed by determining whether cancer cells are benign or malignant. For the handwritten digits the perceptron produced an accuracy of 93.75%, versus 98.75% that can be achieved with an electronic digital neural network. For the tissue biopsy data classification for cancer cells (Figure 8), individual cell nuclei were extracted from breast mass tissue via fine needle aspirate and then imaged with a microscope. These images were previously characterized to distinguish 30 different features including texture, perimeter, radius, etc.. For our experiments, the data for 521 cell nuclei were used for pre-training the network, with a further 75 used as the basis for the testing diagnosis. This follows a very similar process to that used for the handwritten digit tests discussed above. We obtained an 86.67% accuracy versus 98.67% that can be achieved with a digital electronic neural network.

In our experiments we used Intel's approach of evaluating digital microprocessors [98]. Since our system is rather more complex in that it uses input data and weight vectors for the MAC calculations that come from different sources that are multiplexed in time and wavelength, we define the throughput speed based on the temporal data sequence of the electronic output port, in order to be unambiguous. According to the protocol of broadcast-and-delay, each computation cycle consists of one vector dot product between the 49 symbol data and the weighted vectors, resulting in a time data sequence having a length of 48+1+48 symbols, yielding a total duration time of 97 × 84ps. The 49$^{th}$ symbol represents the desired result – ie., the vector dot product resulting from 49 MAC operations, and hence the perceptron throughput is given by 49 / (84ps × 97) = 5.95 Giga-MACs/s. Since each MAC operation consists of two operations — a multiply followed by an accumulate operation —our throughput measured in operations (OPS) is twice that measured in MACs/s, or (49×2)/(84 ps×97) = 11.9 Giga-OPS.

The input data sequence contained 8-bit symbols of 256 discrete levels, reflecting the pixel values of the grey scale image. The 8 bits was limited by our electronic arbitrary waveform generator's intensity resolution. The Waveshaper had a range in attenuation of 35 dB, which is equivalent to a resolution of 11 bits or 33 dB ($=10\times\log_{10}[2^{11}]$). Therefore, every computing cycle had an effective throughput bit rate of (49×2) × 8 / (84 ps × 97) = 95.2 Gigabits/s. For analogue systems such as ours, both the intensity resolution and the bit rate are limited by the system SNR (signal-to-noise ratio). Therefore, in order to have a full resolution of 8-bits, our system needed to have a SNR greater than 20·log10(28) = 48 dB in terms of electric power. This is well within the capability of analogue photonic microwave links, such as the perceptron system that we reported here which had an OSNR > 28 dB.

Our perceptron is the fastest optically based neuromorphic processor ever reported, although making direct comparisons with all of the different approaches is challenging since they vary so widely. As an example, on the one hand systems based on static or continuous sources that perform one-off or single-shot measurements [11, 17, 24] can have a very low latency. However, on the other hand, they also suffer from an extremely low throughput since the input data cannot be in any rapid manner. While our perceptron did have a relatively large latency of ~64 μs, this was purely due to the dispersive delay component which in our case was a simple spool of optical fibre. This did not, however, have any effect on the speed or throughput of our system. Moreover, in fact this can be dramatically reduced or virtually eliminated – easily to less than 200 ps – just by using any type of compact device that can replace the dispersive delay of the fibre, such as sampled Bragg gratings or etalon based tuneable dispersion compensators [99-103] and other approaches [104-107].

## VII. Speed Calculation

Following our definition of throughput and latency introduced in the manuscript, the overall throughput of the deep ONN is roughly the product of each hidden layer's speed and the number of hidden layers, although we note that rigorous and accurate calculation of the throughput is only possible with specific configurations of the network.

Here is a simple example of calculation (this example is just to show the calculations of throughput and latency, the actual performance in terms of prediction accuracies is not the focus of our discussion here): the input waveform/layer is the same as the demonstrated perceptron (49×1 vector at 11.9 Giga Baud with 8-bit resolution, $\tau$= 84ps), the network has a hidden layer that each has 7 fully connected neurons, and an output layer that has 10 fully connected neurons (to match with the number of categories for digits from 0 to 9). As a result, 343 (49×7) and 70 (7×10) wavelengths would be needed in the hidden and output layer, respectively. This can be achieved by using smaller FSR microcombs such as 25GHz across the wide optical band (the C + L bands already reach >11THz wide).

In the hidden layer, each initial electrical output waveform (right after the photodetection and before the digital signal processing) corresponds to the output of a single neuron and has a duration of (49×2−1)×84ps=8.148 ns. Only one time slot of each group of symbols represents the result of matrix multiplication between the input vector and the weight vector that constitutes of 49×2=98 floating point operations. As a result, the throughput of each neuron is given as 98/8.148=12.0275 Giga-OPS. Since different neurons are multiplexed in both the spatial and wavelength domain and detected in parallel, the total throughput of the hidden layer would be 12.0275×7=84.1925 Giga-OPS.

In the output layer, the generated electrical waveform of each neuron has a duration of (7×2−1)×84ps=1.092 ns. Only one time slot of each group of symbols represents the result of matrix multiplication between the input vector (sampled and re-multiplexed waveform from the hidden layer) and the weight vector that constitutes of 7×2=14 floating point operations, thus the throughput would be 14/1.092=12.8205 Giga-FLOPS for each neuron and the total throughput of output layer would be 12.8205×10=128.205 Giga-FLOPS. As such, the total peak throughput of the network would be 84.1925+128.205 =212.3975 Giga-FLOPS. In addition, the latency of the overall network is the sum of each layer's latency, which mainly comes from the dispersive optical buffer and the electrical sampling and multiplexing module. We assume the latency to be 200 ps for the buffer in integrated forms and to be twice of the waveform duration for the re-sampling unit (2×8.148 ns and 2×1.092 ns for the hidden and output layer, respectively), the total latency of the example network would roughly be 18.68 ns. We note that the latency is just a very rough estimation showing how to calculate or measure the performance of our approach, the practical calculations of the latency are subject to more detailed parameters.

The speed of the network has the potential to reach 10 Tera-OPS [12], determined as follows. With 20 layers, each layer featuring 20 neurons and a modulation rate of 25 Giga baud, the overall throughput should be around 20×20×25=10 tera-FLOPS, according to the discussion in the above section. With 8-bit resolution, the total potential throughput in terms of bit rate could reach 10×8=80 Tbps. We note that other widely used techniques in telecommunications such as polarization multiplexing and coherent modulation formats could also potentially boost the computing speed of the proposed neuron network in this work.

Table 1 shows the performance matrices of state-of-art ONNs. We note that it is difficult to directly compare different kinds of ONNs, since on one hand, there are no universal and specific definitions of ONN's parameters. On the other hand, the operation principles of existing ONNs are quite different and have their unique advantages. As such, here we highlight the decent advances of existing works and focus on the speed parameters, including the *latency* and *throughput*, to reflect our ONN's advantages in this aspect.

**Table 1** Performance comparison of state-of-the-art ONNs

| Approach\Parameter | | Compatibility with digital electronics | Latency | Throughput speed per unit | |
|---|---|---|---|---|---|
| | | | | OPS | bits/s |
| Diffraction devices [17] | | — | < 10 ns | — | — |
| Integrated couplers [3] | | — | < 0.1 ns | — | — |
| Reservoir computing [20] | | Yes | < 1 μs | 17.6 G | — |
| Spike computing [23] | | Yes | < 1 μs | 8 G | 8 G |
| Spike computing [24] | | — | < 0.1 μs | — | — |
| [11] | Single Perceptron | Yes | 64 μs | 11.9 G | 95.2 G |
| [12] | Deep ONN | Yes | >18.68 ns | >10 T | >80 T |

"—" denotes the corresponding parameter is either not demonstrated or not indicated in the work.

## VIII. CONCLUSIONS

We report an optical neural network consisting of a single perceptron that operates with an integrated optical Kerr micro-comb source. The system achieves a single processor throughput speed of 11.9 Giga-OPS/s, equivalent to 95.2 Gigabits/s. We demonstrate benchmark tests including cancer cell diagnosis and handwritten digit recognition. We outline different approaches to scale the network to deep learning ONN architectures that have significantly increased processing power and throughput speed. This is possible because of the high level of parallelism that can be realized via simultaneous time, spatial, and wavelength multiplexing. Our approach has significant possibilities for real-time analysis of high dimensional data for advanced applications.

Competing interests: The authors declare no competing interests.